\documentclass{icrc2009}

\usepackage{graphicx}   % for including figures
\usepackage[caption=false]{caption}    % for captions
\usepackage[font=footnotesize]{subfig} % subfig.sty for a double column floating figure using two subfigures
\usepackage{fixltx2e}
\usepackage{url}

\newcommand{\shorttitle}[1]%
{\markboth{Proceedings of the 31\MakeLowercase{$^{st}$} ICRC, {\L}\'{o}d\'{z} 2009}{#1} }
\newcommand{\etal}{\MakeLowercase{\textit{et al. }}} % "et al."

\begin{document}
\title{Large-scale sidereal anisotropy of multi-TeV galactic cosmic rays and the heliosphere}

\author{
\IEEEauthorblockN{
M.~Amenomori\IEEEauthorrefmark{1}, X.~J.~Bi\IEEEauthorrefmark{2},
D.~Chen\IEEEauthorrefmark{3}, S.~W.~Cui\IEEEauthorrefmark{4},
Danzengluobu\IEEEauthorrefmark{5}, L.~K.~Ding\IEEEauthorrefmark{2},\\
X.~H.~Ding\IEEEauthorrefmark{5}, C.~Fan\IEEEauthorrefmark{6},
C.~F.~Feng\IEEEauthorrefmark{6}, Zhaoyang Feng\IEEEauthorrefmark{2},
Z.~Y.~Feng\IEEEauthorrefmark{7}, X.~Y.~Gao\IEEEauthorrefmark{8},
Q.~X.~Geng\IEEEauthorrefmark{8},\\
Q.~B.~Gou\IEEEauthorrefmark{2}, H.~W.~Guo\IEEEauthorrefmark{5},
H.~H.~He\IEEEauthorrefmark{2}, M.~He\IEEEauthorrefmark{6},
K.~Hibino\IEEEauthorrefmark{9}, N.~Hotta\IEEEauthorrefmark{10},
Haibing~Hu\IEEEauthorrefmark{5}, H.~B.~Hu\IEEEauthorrefmark{2},\\
J.~Huang\IEEEauthorrefmark{2}, Q.~Huang\IEEEauthorrefmark{7},
H.~Y.~Jia\IEEEauthorrefmark{7}, L.~Jiang\IEEEauthorrefmark{8}\,\IEEEauthorrefmark{2},
F.~Kajino\IEEEauthorrefmark{11}, K.~Kasahara\IEEEauthorrefmark{12},
Y.~Katayose\IEEEauthorrefmark{13},\\
C.~Kato\IEEEauthorrefmark{14}, K.~Kawata\IEEEauthorrefmark{3},
Labaciren\IEEEauthorrefmark{5}, G.~M.~Le\IEEEauthorrefmark{15},
A.~F.~Li\IEEEauthorrefmark{6}, H.~C.~Li\IEEEauthorrefmark{4}\,\IEEEauthorrefmark{2}, 
J.~Y.~Li\IEEEauthorrefmark{6}, C.~Liu\IEEEauthorrefmark{2},\\
Y.-Q.~Lou\IEEEauthorrefmark{16}, H.~Lu\IEEEauthorrefmark{2},
X.~R.~Meng\IEEEauthorrefmark{5}, K.~Mizutani\IEEEauthorrefmark{12}\,\IEEEauthorrefmark{17},
J.~Mu\IEEEauthorrefmark{8}, K.~Munakata\IEEEauthorrefmark{14},
A.~Nagai\IEEEauthorrefmark{18},\\
H.~Nanjo\IEEEauthorrefmark{1}, M.~Nishizawa\IEEEauthorrefmark{19},
M.~Ohnishi\IEEEauthorrefmark{3}, I.~Ohta\IEEEauthorrefmark{20},
S.~Ozawa\IEEEauthorrefmark{12}, T.~Saito\IEEEauthorrefmark{21},
T.~Y.~Saito\IEEEauthorrefmark{22},\\
M.~Sakata\IEEEauthorrefmark{11}, T.~K.~Sako\IEEEauthorrefmark{3},
M.~Shibata\IEEEauthorrefmark{13}, A.~Shiomi\IEEEauthorrefmark{23},
T.~Shirai\IEEEauthorrefmark{9}, H.~Sugimoto\IEEEauthorrefmark{24},
M.~Takita\IEEEauthorrefmark{3},\\
Y.~H.~Tan\IEEEauthorrefmark{2}, N.~Tateyama\IEEEauthorrefmark{9},
S.~Torii\IEEEauthorrefmark{12}, H.~Tsuchiya\IEEEauthorrefmark{25},
S.~Udo\IEEEauthorrefmark{9}, B.~Wang\IEEEauthorrefmark{2},
H.~Wang\IEEEauthorrefmark{2}, Y.~Wang\IEEEauthorrefmark{2},\\
Y.~G.~Wang\IEEEauthorrefmark{6}, H.~R.~Wu\IEEEauthorrefmark{2},
L.~Xue\IEEEauthorrefmark{6}, Y.~Yamamoto\IEEEauthorrefmark{11},
C.~T.~Yan\IEEEauthorrefmark{26}, X.~C.~Yang\IEEEauthorrefmark{8},
S.~Yasue\IEEEauthorrefmark{27},\\
Z.~H.~Ye\IEEEauthorrefmark{28}, G.~C.~Yu\IEEEauthorrefmark{7},
A.~F.~Yuan\IEEEauthorrefmark{5}, T.~Yuda\IEEEauthorrefmark{9},
H.~M.~Zhang\IEEEauthorrefmark{2}, J.~L.~Zhang\IEEEauthorrefmark{2},
N.~J.~Zhang\IEEEauthorrefmark{6},\\
X.~Y.~Zhang\IEEEauthorrefmark{6}, Y.~Zhang\IEEEauthorrefmark{2},
Yi~Zhang\IEEEauthorrefmark{2}, Ying~Zhang\IEEEauthorrefmark{7}\,\IEEEauthorrefmark{2},
Zhaxisangzhu\IEEEauthorrefmark{5} and X.~X.~Zhou\IEEEauthorrefmark{7}\\
(The Tibet AS$\gamma$ Collaboration)} and J.~K\'ota\IEEEauthorrefmark{29}\\ 
\\
\IEEEauthorblockA{\IEEEauthorrefmark{1}Department of Physics, Hirosaki University, Hirosaki 036-8561, Japan.}
\IEEEauthorblockA{\IEEEauthorrefmark{2}Key Laboratory of Particle Astrophysics, Institute of High Energy Physics, Chinese Academy of Sciences,\\ Beijing 100049, China.}
\IEEEauthorblockA{\IEEEauthorrefmark{3}Institute for Cosmic Ray Research, University of Tokyo, Kashiwa 277-8582, Japan.}
\IEEEauthorblockA{\IEEEauthorrefmark{4}Department of Physics, Hebei Normal University, Shijiazhuang 050016, China.}
\IEEEauthorblockA{\IEEEauthorrefmark{5}Department of Mathematics and Physics, Tibet University, Lhasa 850000, China.}
\IEEEauthorblockA{\IEEEauthorrefmark{6}Department of Physics, Shandong University, Jinan 250100, China.}
\IEEEauthorblockA{\IEEEauthorrefmark{7}Institute of Modern Physics, SouthWest Jiaotong University, Chengdu 610031, China.}
\IEEEauthorblockA{\IEEEauthorrefmark{8}Department of Physics, Yunnan University, Kunming 650091, China.}
\IEEEauthorblockA{\IEEEauthorrefmark{9}Faculty of Engineering, Kanagawa University, Yokohama 221-8686, Japan.}
\IEEEauthorblockA{\IEEEauthorrefmark{10}Faculty of Education, Utsunomiya University, Utsunomiya 321-8505, Japan.}
\IEEEauthorblockA{\IEEEauthorrefmark{11}Department of Physics, Konan University, Kobe 658-8501, Japan.}
\IEEEauthorblockA{\IEEEauthorrefmark{12}Research Institute for Science and Engineering, Waseda University, Tokyo 169-8555, Japan.}
\IEEEauthorblockA{\IEEEauthorrefmark{13}Faculty of Engineering, Yokohama National University, Yokohama 240-8501, Japan.}
\IEEEauthorblockA{\IEEEauthorrefmark{14}Department of Physics, Shinshu University, Matsumoto 390-8621, Japan.}
\IEEEauthorblockA{\IEEEauthorrefmark{15}National Center for Space Weather, China Meteorological Administration, Beijing 100081, China.}
\IEEEauthorblockA{\IEEEauthorrefmark{16}Physics Department and Tsinghua Center for Astrophysics, Tsinghua University, Beijing 100084, China.}
\IEEEauthorblockA{\IEEEauthorrefmark{17}Saitama University, Saitama 338-8570, Japan.}
\IEEEauthorblockA{\IEEEauthorrefmark{18}Advanced Media Network Center, Utsunomiya University, Utsunomiya 321-8585, Japan.}
\IEEEauthorblockA{\IEEEauthorrefmark{19}National Institute of Informatics, Tokyo 101-8430, Japan.}
\IEEEauthorblockA{\IEEEauthorrefmark{20}Sakushin Gakuin University, Utsunomiya 321-3295, Japan.}
\IEEEauthorblockA{\IEEEauthorrefmark{21}Tokyo Metropolitan College of Industrial Technology, Tokyo 116-8523, Japan.}
\IEEEauthorblockA{\IEEEauthorrefmark{22}Max-Planck-Institut f\"ur Physik, M\"unchen D-80805, Deutschland.}
\IEEEauthorblockA{\IEEEauthorrefmark{23}College of Industrial Technology, Nihon University, Narashino 275-8576, Japan.}
\IEEEauthorblockA{\IEEEauthorrefmark{24}Shonan Institute of Technology, Fujisawa 251-8511, Japan.}
\IEEEauthorblockA{\IEEEauthorrefmark{25}RIKEN, Wako 351-0198, Japan.}
\IEEEauthorblockA{\IEEEauthorrefmark{26}Institute of Disaster Prevention Science and Technology, Yanjiao 065201, China.}
\IEEEauthorblockA{\IEEEauthorrefmark{27}School of General Education, Shinshu University, Matsumoto 390-8621, Japan.}
\IEEEauthorblockA{\IEEEauthorrefmark{28}Center of Space Science and Application Research, Chinese Academy of Sciences, Beijing 100080, China.}
\IEEEauthorblockA{\IEEEauthorrefmark{29}Lunar and Planetary Laboratory, University of Arizona, Tucson, AZ 87721, USA.}}
% please write the preseter's name and short title (3-4 words maximum)
%    which will appear at the header of the even pages.
\shorttitle{K. Munakata \etal Sidereal anisotropy and the heliosphere}
\maketitle

\begin{abstract}

We develop a model anisotropy best-fitting to the two-dimensional sky-map of multi-TeV galactic cosmic ray (GCR) intensity observed with the Tibet III air shower (AS) array. By incorporating a pair of intensity excesses in the hydrogen deflection plane (HDP) suggested by Gurnett et al., together with the uni-directional and bi-directional flows for reproducing the observed global feature, this model successfully reproduces the observed sky-map including the ``skewed'' feature of the excess intensity from the heliotail direction, whose physical origin has long remained unknown. These additional excesses are modeled by a pair of the northern and southern Gaussian distributions, each placed $\sim$50$^{\circ}$ away from the heliotail direction. The amplitude of the southern excess is as large as $\sim$0.2 \%, more than twice the amplitude of the northern excess. This implies that the Tibet AS experiment discovered for the first time a clear evidence of the significant modulation of GCR intensity in the heliotail and the asymmetric heliosphere.
\end{abstract}


\section{Introduction}

\begin{figure*}[th]
  \centering
  \includegraphics[width=\linewidth]{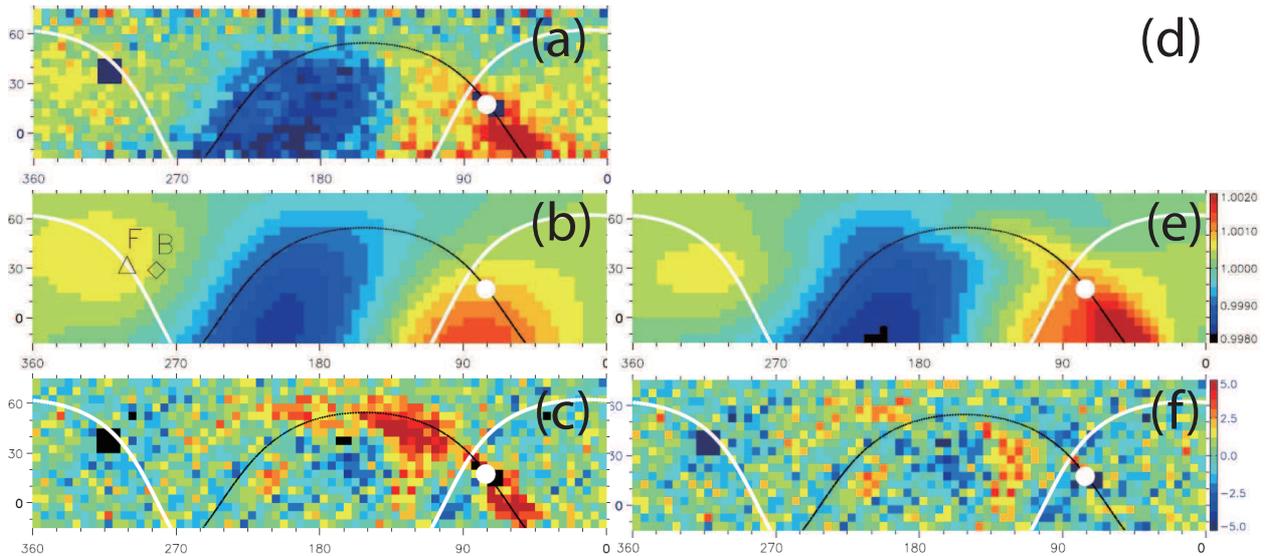}
  \caption{2D-sky maps of the observed and reproduced GCR intensity. The panels display the normalized GCR intensity or significance in $5^{\circ}\times5^{\circ}$ pixels in a color-coded format as a function of the right ascension on the horizontal axis and the declination on the vertical axis. In this figure, the average intensity in each declination belt is normalized to unity. These sky-maps cover $360^{\circ}$ of the right ascension but cover only $90^{\circ}$ of the declination due to the event selection criterion limiting zenith angles to $\le 45^{\circ}$ (see text). The data in 16 pixels containing the known and possible gamma ray sources are excluded from the best-fit calculation and indicated by black color. The white curve indicates the galactic plane, while the black curve displays the HDP plane suggested by Gurnett et al. \cite{paper07}, which is calculated as a plane normal to the orientation of $\alpha=332.1^{\circ}$ and $\delta=35.5^{\circ}$. In each panel, the heliotail direction  ($\alpha=75.9^{\circ}$ and $\delta=17.4^{\circ}$) is indicated by a white solid circle in the HDP plane. Each panel displays, (a): the observed intensity ($i_{n,m}^{obs}$), (b): the best-fit component anisotropy ($i_{n,m}^{GA}$) reproducing the global anisotropy (GA), (c): the significance of the residual anisotropy remaining after the subtraction of $i_{n,m}^{GA}$ from $i_{n,m}^{obs}$ ( $(i_{n,m}^{obs}-i_{n,m}^{GA})/ \sigma_{n,m}$), (d): the best-fit component anisotropy ($i_{n,m}^{AE}$) reproducing the additional excess (AE) intensity, (e): the total model anisotropy $i_{n,m}$ best-fitting to (a), (f): the residual significance ($(i_{n,m}^{obs}-i_{n,m})/\sigma_{n,m}$). In panel (b), the open triangle with an attached character ``F" indicates the LISMF orientation by Frisch ($\alpha=300.9^{\circ}$ and $\delta=32.2^{\circ}$) \cite{paper12}, while the open diamond with ``B" indicates the orientation of the best-fit BDF (see text and Table I).}
  \label{simp_fig}
\end{figure*}

\begin{table*}[th]
  \caption{Best-fit parameters in $I_{n,m}^{GA}$ and $I_{n,m}^{AE}$ in eqs. (3) and (4). Note that $\delta_2$ in the upper table is derived according to our definition that the reference axes $(\alpha_1,\delta_1)$ and $(\alpha_2,\delta_2)$ are perpendicular to each other \cite{paper02}.}
  \label{table_simple}
  \centering
  \begin{tabular}{cccccccccccc}
    \hline \hline 
    $a_{1\perp}(\%)$& $a_{1\parallel}(\%)$&$a_{2\parallel}(\%)$&$\alpha_1(^{\circ})$&$\delta_1(^{\circ})$&$\alpha_2(^{\circ})$&$\delta_2(^{\circ})$&$b_1(\%)$&$b_2(\%)$&$\sigma_{\parallel}(^{\circ})$&$\sigma_{\perp}(^{\circ})$&$ \Phi(^{\circ})$\\
    \hline 
    0.141 & 0.006 & 0.140 & 37.5 & 37.5 & 102.5 & -28.9 & 0.234 & 0.100 & 25.0 & 10.0 & 52.5\\
    \hline
  \end{tabular}
\end{table*}

The Tibet III Air Shower Array experiment has been conducted at Yangbajing (90.522$^{\circ}$E, 30.102$^{\circ}$N; 4300 m above sea level) in Tibet, China. The array composed of 533 scintillation counters of 0.5 m$^2$ each covers a detection area of 22,050 m$^2$ achieving a trigger rate of $\sim$680 Hz. GCR events are selected for analyses, if all the following criteria are met: (i) any four-fold coincidence occurs in the counters with each recording more than 0.8 particles in charge, (ii) the air shower core position is located in the array, (iii) the zenith angle of arrival direction is $\le 45^{\circ}$. With all these criteria, the array has the modal GCR energy of $\sim$5 TeV. The angular resolution of the arrival direction of each shower is estimated to be $\sim$0.9$^{\circ}$ from Monte Carlo simulations, and this was also verified by measuring the Moon's shadow in GCRs \cite{paper01}. In this paper, we analyze a total of 37 billion air shower events recorded in 1318.9 live days from November 1999 to October 2005.

Fig. 1(a) shows the observed GCR intensity in $5^{\circ}\times5^{\circ}$ pixels in a color-coded format as a function of the right ascension ($\alpha$) on the horizontal axis and the declination ($\delta$) on the vertical axis. For producing this figure, we first obtain the GCR intensity $I_{n,m}^{obs}$ in $n$-th right ascension and $m$-th declination pixel. We then normalize the average of $I_{n,m}^{obs}$ in each declination belt to unity and get the normalized model intensity $i_{n,m}^{obs}$ plotted in this figure. This sky-map covers 360$^{\circ}$ of $\alpha$ but covers only 90$^{\circ}$ of $\delta$ ranging from -15$^{\circ}$ to +75$^{\circ}$ due to the event selection criterion limiting zenith angles to  $\le 45^{\circ}$. The map clearly shows a significant anisotropy, consisting of a $\sim$0.2 \% excess at $\alpha\sim60^{\circ}$ and $\delta\sim-10^{\circ}$ and a $\sim$0.2 \% deficit at $\alpha\sim180^{\circ}$ and $\delta\sim0^{\circ}$, each observed with a statistical significance more than ten times the statistical error. There is also the ``skewed" feature seen in the region of excess intensity as pointed by \cite{paper02}. In the next section, we develop a model anisotropy reproducing this observed global feature as well as the ``skewed" feature.

\section{Analysis and result}

We develop a model anisotropy $I_{n,m}$ consisting of two components as
\begin{equation}
I_{n,m}=I_{n,m}^{GA}+I_{n,m}^{AE}
\end{equation}
where $I_{n,m}^{GA}$ and $I_{n,m}^{AE}$ respectively denote the global anisotropy (GA) and the additional excess (AE) intensity as described below. We first normalize the average of $I_{n,m}$ in each declination belt to unity and get the normalized model intensity $i_{n,m}=i_{n,m}^{GA}+i_{n,m}^{AE} $, the same way as we did for the observed data to produce Fig. 1(a). By comparing $i_{n,m}$ with $i_{n,m}^{obs}$ in Fig. 1(a), we obtain best-fit parameters minimizing the residual $S$ defined, as
\begin{equation}
S=\Sigma_{n=1}^{72}\Sigma_{m=1}^{18}(i_{n,m}^{obs}-i_{n,m})^2/\sigma_{n,m}^2
\end{equation}
where $\sigma_{n,m}$ is the statistical error of the intensity in $(n,m)$ pixel. From this best-fitting, we excluded 16 pixels containing the known and possible gamma ray sources. These pixels also include the ``region A" reported by Milagro experiment \cite{paper03}. 

The observed angular separation between the excess and the deficit in Fig. 1(a) is only  $\sim120^{\circ}$, which is much smaller than 180$^{\circ}$ expected from a uni-directional flow (UDF) but significantly larger than 90$^{\circ}$ expected from a bi-directional flow (BDF). Only a combination of the uni-directional and bi-directional flows can achieve a reasonable fit to the global feature of the observed anisotropy. From this point of view, we perform a best-fit calculation to the observed global anisotropy (GA) with a model intensity $I_{n,m}^{GA}$ expressed, as
\begin{eqnarray}
\lefteqn{I_{n,m}^{GA}=a_{1\perp}\cos\chi_1(n,m:\alpha_1,\delta_1)} \nonumber\\
&&          +a_{1\parallel}\cos\chi_2(n,m:\alpha_2,\delta_2)\nonumber\\
&&          +a_{2\parallel}\cos^2\chi_2(n,m:\alpha_2,\delta_2)
\end{eqnarray}
where $a_{1\perp}$ and $a_{1\parallel}$ are amplitudes of UDFs perpendicular and parallel to the BDF, respectively, $a_{2\parallel}$ is the amplitude of the BDF, $(\alpha_1,\delta_1)$ and $(\alpha_2,\delta_2)$ are respectively right ascensions and declinations of the reference axes of the perpendicular UDF and BDF and $\chi_1$ ($\chi_2$) is the angle of the center of ($n, m$) pixel measured from the reference axis of the perpendicular UDF (BDF) \cite{paper02}. Fig. 1(b) displays $i_{n,m}^{GA}$ derived from best-fitting to Fig. 1(a). While the model anisotropy in eq. (3) well reproduces the observed global feature in Fig. 1(a), it is obviously too simple for reproducing the observed feature of the excess intensity at around $\alpha\sim60^{\circ}$ and $\delta\sim-10^{\circ}$. This is demonstrated clearly in Fig. 1(c) displaying the significance of the residual anisotropy remaining after the subtraction of $i_{n,m}^{GA}$ from Fig. 1(a).

It is clear that the observed anisotropy contains an additional excess intensity along a plane which is almost perpendicular to the galactic plane. This additional excess intensity was first reported as the ``skewed" feature from the long-term observations of sub-TeV GCR intensity with underground muon detectors, but its physical origin has remained unknown \cite{paper05}\cite{paper06}. It is also clear in Fig. 1(c) that the additional excess intensity extends along Gurnett's HDP displayed by a black curve \cite{paper07}. We model this additional excess (AE) intensity by a pair of Gaussians placed in Gurnett's HDP, each centered away from the heliotail direction by an angle $\Phi$, as
\begin{eqnarray}
\lefteqn{I\!_{\!n,m}^{\!AE}\!=\!\big\{b_1\exp(-\frac{(\phi_{n,m}-\Phi)^2}{2\sigma_{\phi}^2})} \nonumber \\
&&+b_2\exp(-\frac{(\phi_{n,m}+\Phi)^2}{2\sigma_{\phi}^2})\big\}\exp(-\frac{\theta_{n,m}^2}{2\sigma_{\theta}^2})
\end{eqnarray}
where $b_1$ and $b_2$ are amplitudes, $\sigma_{\phi}$ and $\sigma_{\theta}$ are widths parallel and perpendicular to the HDP respectively, $\phi_{n,m}$ is the ``longitude" of the center of ($n, m$) pixel measured from the heliotail along the HDP and $\theta_{n,m}$ is the ``latitude" measured from the HDP. Fig. 1(d) displays the best-fit $i_{n,m}^{AE}$, while Fig. 1(e) displays the combined model anisotropy $i_{n,m}$ best-fitting to $i_{n,m}^{obs}$ in Fig. 1(a). As seen in Fig. 1(f) showing the significance of the residual anisotropy remaining after the subtraction of $i_{n,m}$ from Fig. 1(a), this new model $i_{n,m}$ successfully reproduces the observed anisotropy including the ``skewed" feature of the excess intensity as well as the global feature. Eleven best-fit parameters ( $a_{1\perp}, a_{1\parallel}, a_{2\parallel}, \alpha_1, \delta_1, \alpha_2, b_1, b_2, \sigma_{\phi}, \sigma_{\theta}, \Phi$) minimizing the residual $S$ in eq. (2) are listed in Table I. The minimum $S$ divided by the degree of freedom (1269=72$\times$18-16-11) is 1.791. Note that five amplitudes ($a_{1\perp}, a_{1\parallel}, a_{2\parallel}, b_1, b_2$) are uniquely determined by the least-square technique for each set of remaining six angle parameters ($\alpha_1, \delta_1, \alpha_2, \sigma_{\phi}, \sigma_{\theta}, \Phi$).

\section{Summary and discussion}

The Larmor radius of 5 TeV GCR protons in a 3 $\mu$G interstellar magnetic field is $r_L \sim$0.002 pc or 400 AU, which is comparable to the scale of the heliosphere in the nose direction toward the upstream of the interstellar wind. This is one reason why the heliospheric modulation of the GCR intensity has been considered to be negligible in this energy region. The heliosphere, however, is also known to have a long heliotail extending over thousands of AU, much longer than  $r_L$ of 5 TeV GCRs \cite{paper08}. The GCR modulation in the heliotail still remains possible, although it is not fully understood yet.

Recent observations also have suggested asymmetries of the heliosphere. Based on the deflection of the interstellar neutral hydrogen flow vector from the helium flow vector observed at 1 AU, Lallement et al. deduced the Hydrogen Deflection Plane (HDP) containing the interstellar wind velocity and the Local Interstellar Magnetic Field (LISMF) and suggested a possible north-south asymmetry of the heliosphere due to the magnetic pressure of the LISMF \cite{paper09}. From the observation of 2-3 kHz radio emission from the outer heliosphere, Gurnett et al. \cite{paper07} also deduced the HDP almost perpendicular to the galactic plane in a reasonable agreement with Lallement et al. Opher et al. analyzed the difference in the heliocentric distances to the solar wind termination shock observed by $\it{Voyagers 1}$ and $\it{2}$ at different heliographic latitudes and longitudes. By comparing the difference with the Magneto-Hydro-Dymanic (MHD) simulations, they derived the north-south and east-west asymmetries of the heliosphere \cite{paper10}\cite{paper11}.

To achieve a good fit to the observed sky-map by Tibet AS experiment, the model anisotropy is required to include an additional excess intensity in the Gurnett's HDP plane. This additional anisotropy is best modeled with a pair of Gaussians, each centerd $\sim50^{\circ}$ away form the heliotail direction (see $\Phi$ in Table I). The amplitude of the southern excess ($b_1$) is as large as $\sim$0.2 \% more than twice the amplitude of the northern excess ($b_2$), possibly indicating a significant north-south asymmetry of the heliosphere. The Tibet AS experiment succeeded for the first time to reveal a clear signature of the asymmetric heliosphere in the sidereal anisotropy of the multi-TeV GCR intensity.

\newpage
%\noindent
\section{Acknowledgments}

The collaborative experiment of the Tibet Air Shower Arrays has been performed under the auspices of the Ministry of Science and Technology of China and the Ministry of Foreign Affairs of Japan. This work was supported in part by Grant-in-Aid for Scientific Research on Priority Areas from the Ministry of Education, Culture, Sports, Science and Technology, by Grants-in-Aid for Science Research from the Japan Society for the Promotion of Science in Japan, and by the Grants-in-Aid from the National Natural Science Foundation of China and the Chinese Academy of Sciences.


\vspace{6.5 mm}
\begin{thebibliography}{99}
\bibitem{paper01} Amenomori et al., M., 1993, Phys. Rev. D, {\bf 47}, 2675-2681.
\bibitem{paper02} Amenomori et al., M., 2007, AIP Conf. Proc., {\bf 932}, 283-289.
\bibitem{paper03} Abdo, A. A. et al., 2008, Phys. Rev. Lett., {\bf 101}, 221101-1-221101-4.
\bibitem{paper04} Mizoguchi, Y. et al., 2009, Proc. 31st ICRC, Lodz, SH3.2, icrc0388.
\bibitem{paper05} Hall, D. et al., 1998, J. Geophys. Res., {\bf 103}, 367-372.
\bibitem{paper06} Hall, D. et al., 1999, J. Geophys. Res., {\bf 104}, 6737-6749.
\bibitem{paper07} Gurnett, D. A. et al., 2006, AIP Conf. Proc., {\bf 858}, 129-134.
\bibitem{paper08} Washimi, H., and Tanaka, T., 1996, Space Sci. Rev., {\bf 78}, 85-94.
\bibitem{paper09} Lallement, R. et al., 2005, Science, {\bf 307}, 1447-1449.
\bibitem{paper10} Opher, M., Stone, E. C., and Liewer, P. C., 2006, Astrophys. J., {\bf 640}, L71-L74.
\bibitem{paper11} Opher, M., Stone, E. C., and Gombosi, T. I., 2007, Science, {\bf 316}, 875-878.
\bibitem{paper12} Frisch, P.,C., 1996, Space Sci. Rev., {\bf 78}, 213-222
\end{thebibliography}
\end{document}